\documentstyle[aps,multicol]{revtex}
\begin{document}
\title{Three Ignored Densities, Frame-Independent\\ 
Thermodynamics and Broken Galilean Symmetry}
\author{Peter Kost\"adt and Mario Liu~\cite{email}}
\address{Institut f\"ur Theoretische Physik, Universit\"at Hannover,
      30167 Hannover, Germany, EC}
\date{\today}\maketitle
\begin{abstract}
A system's invariance under Galilean transformation implies three locally 
conserved densities. Including them as variables, the thermodynamics 
is rendered explicitly frame independent, dissipative mass currents are 
shown to vanish, and spontaneously broken Galilean symmetry 
becomes a sensible concept in condensed systems.   
\end{abstract}
\draft\vspace{-.5cm}\begin{multicols}{2}

\section{Introduction}
The invariance under Galilean boosts of a closed system implies the 
conservation of 

\begin{equation}\label{1}
{\bf B}=M{\bf R}_0,
\end{equation}
where $M$ is the system's mass, and ${\bf R}_0$ its center of mass 
coordinate at the time $t=0$~\cite{Doug}. Despite the appearance of an 
initial value that can always be set to zero by an appropriate choice of the 
coordinate, {\bf B} is an additive and  locally conserved quantity, in 
complete analogy to energy, mass, momentum and angular momentum. As these 
are all thermo- and hydrodynamic variables, it is a valid question why {\bf B}
is never included. This paper shows the usefulness, even necessity, of 
including {\bf B}, summarized in the following list.

{\bf i)} The inclusion of {\bf B} as a variable is a necessary condition for 
the formulation of a frame independent thermodynamic theory: Starting from 
the rest frame expression, say $dE=TdS$, it may seem all right to add ${\bf 
V}\!\cdot d{\bf G}$ for a boosted system with the momentum ${\bf G}$, and 
add ${\bf\Omega}\cdot d{\bf L}$ if it also rotates with the angular  
momentum ${\bf L}$, but this is not enough. As shown in section~\ref{2}, 
this is only correct for frames in which ${\bf \Omega\,\|\,G}$, while 
generally the energy $E$ also depends on ${\bf B}$. 

Before going on with the list, it is convenient to introduce a name 
for 
\begin{equation}
{\bf B}=M({\bf R}-{\bf\dot R}\,t)={\textstyle\int}{\rm d}^3\!x\,
        (\varrho\,\mbox{\boldmath$x$}-\mbox{\boldmath$g$}\,t)\,,
\label{comic}\end{equation}
where ${\bf R}={\bf R}_0+{\bf\dot R}t$ is the time dependent center of mass 
coordinate, with a constant time derivative ${\bf\dot R}={\bf G}/M$, while 
$\varrho$ and $\mbox{\boldmath$g$}$ denote the density of mass and momentum, 
respectively.  Following Schwinger~\cite{Schw}, we shall refer to {\bf B} as the
``booster", and to ${\bf b}=\varrho\,\mbox{\boldmath$x$}
-\mbox{\boldmath$g$}\,t$ as the ``booster density". Including the energy's 
dependence on the booster, $dE=\dots+{\bf A}\!\cdot d{\bf B}$, the conjugate 
variable ${\bf A}$ will turn out to be ${\bf\Omega\times\dot R}$. 

{\bf ii)} One of the more direct results of the thermodynamic theory are the 
equilibrium conditions. Again, the correct derivation requires the inclusion 
of  the booster, although this is never done. If one derives them 
following Landau and Lifshitz~\cite{LL5} by maximizing the entropy while 
holding constant: the energy, mass, momentum, and angular momentum, we find 
as equilibrium conditions: constant temperature and chemical potential, with 
the velocity given as $\mbox{\boldmath$v$}={\bf 
V}+{\bf\Omega}\times\mbox{\boldmath$x$}$, where ${\bf V}$ and ${\bf\Omega}$ 
are again constant. Due to the lack of a time dependent term, however, this 
velocity is less general than the usual expression for a rigid body, 
$\mbox{\boldmath$v$}={\bf\dot R}+{\bf\Omega}\times(\mbox{\boldmath$x$}-{\bf 
R})$, or equivalently \begin{equation}
\mbox{\boldmath$v$}=({\bf\dot R-\Omega\times R_0})+{\bf\Omega}\times
\mbox{\boldmath$x$}-({\bf\Omega\times\dot R})\,t\,. 
\label{rig}\end{equation} 
(The difference is in the last term.) This is worrisome as there is no 
reason whatever why a system executing the general motion should not be in 
equilibrium. If, however, the booster is also held constant in the above 
calculation, one does arrive at the general expression, Eq(\ref{rig}). At 
the same time, one finds that the chemical potential, instead of being 
constant, now satisfies 
$\partial_t\mbox{\boldmath$v$}+\mbox{\boldmath$\nabla$}\mu =0$. 
Reassuringly, this is exactly the expression the Navier-Stokes equation 
reduces to for vanishing entropy production, in equilibrium.  

{\bf iii)} Local conservation of the mass density holds if the continuity 
equation, $\partial_t\varrho+ 
\mbox{\boldmath$\nabla$} \!\cdot\mbox{\boldmath$\jmath$}=0$, is satisfied 
--- irrespective what form the mass current $\mbox{\boldmath$\jmath$}$ 
actually assumes. Usually, this form is taken as 
$\mbox{\boldmath$\jmath$}=\varrho\,\mbox{\boldmath$v$}$, although it has 
never been properly deduced; rather, it is accepted as a statement of 
microscopic plausibility, or the summary of countless experiments. So no 
objection was, or could have been, raised, when Dzyaloshinskii and Volovik, 
in their classic paper~\cite{Vol}, propose to include dissipative mass 
currents such as \begin{equation}\label{6}
\mbox{\boldmath$\jmath$} 
-\varrho\,\mbox{\boldmath$v$} \sim\mbox{\boldmath$\nabla$}\mu. 
\end{equation}
(These dissipative terms results from a tempting, even natural, step to 
take when setting up the hydrodynamic equations.) On the other hand, there 
is a well-hidden footnote by Landau and Lifshitz~\cite{LL6} that purports to 
rule out this type of terms --- but actually falls short of being ironclad 
if scrutinized: It invokes the center of mass motion to show that 
$\int\!{\rm d}^3\!x\,\mbox{\boldmath$\jmath$}=\int\!{\rm 
d}^3\!x\,\mbox{\boldmath$g$}$ must prevail, where the integration is to be 
taken over the volume of the system. For reasons of Galilean invariance, the 
momentum density is given as 
$\mbox{\boldmath$g$}=\varrho\,\mbox{\boldmath$v$}$. So a clear-cut proof of 
$\mbox{\boldmath$\jmath$}=\mbox{\boldmath$g$}$ would indeed serve as a sound 
argument for ruling out any dissipative mass currents. Unfortunately, 
$\int\!{\rm d}^3\!x\,\mbox{\boldmath$\jmath$}=\int\!{\rm 
d}^3\!x\,\mbox{\boldmath$g$}$  is less confining, and the reader is left 
wondering about terms that vanish only if integrated. 

As will be discussed in section~\ref{3}, taking the booster-density 
$\varrho\,\mbox{\boldmath$x$} -\mbox{\boldmath$g$}\,t$ as a 
locally conserved quantity that satisfies a continuity equation quickly 
leads to the result $\mbox{\boldmath$\jmath$}=\mbox{\boldmath$g$}$. The 
proof takes place in very much the same way as that deducing the symmetry 
of the stress tensor from the local conservation of angular momentum. 
(This result also has relativistic ramifications, as it shows that out of 
the two versions of relativistic hydrodynamics, by Eckart and by 
Landau-Lifshitz, only the latter is a proper 
generalization~\cite{uniq}. Any linear combination of both is also ruled 
out.) 

{\bf iv)} The angular momentum frequently contains an extensive part, ${\bf 
s}$, that is usually referred to as its intrinsic contribution, 
\begin{equation}
{\bf L}={\textstyle\int}{\rm d}^3\!x\,(\mbox{\boldmath$x$}\!\times\!
             \mbox{\boldmath$g$}+{\bf s}). 
\label{intr}\end{equation}
One example of ${\bf s}$ is the spin density. Due to the close 
relationship between ${\bf L}$ and ${\bf B}$ --- the 
latter being, relativistically, part of the 4-tensor of the angular 
momentum density --- it is certainly not far-fetched to question whether a 
corresponding extensive contribution to the booster may exist, 
\begin{equation}\label{intr2}
{\bf B}={\textstyle\int}{\rm d}^3\!x\,(\varrho\,\mbox{\boldmath$x$}
             -\mbox{\boldmath$g$}\,t+{\bf k}). 
\end{equation}
(Relativistically, the components of the 4-angular-momentum density mix 
under a Lorentz boost. And  ${\bf s}$, the extensive contribution of the 
angular momentum in one frame,  will lead to ${\bf k}$ in others. The more 
precise question is then: Whether ${\bf k}$ exists in the rest frame, same 
as ${\bf s}$.) Microscopically, ${\bf k}$ may be conceived as arising from 
individual mass dipole moments of the molecules or atoms that are not 
compensated by shifting the coordinates of the particles, say because they 
are embedded in a lattice, pivoted off center. 

Although the intrinsic angular momentum ${\bf s}$ has frequently been 
included in hydrodynamic considerations, other authors see this as ad hoc 
and inconsistent, because ${\bf s}$ is a relaxing variable, and its 
inclusion only affects high frequency, non-hydrodynamic phenomena. 
Redefining the momentum density, they then argue that ${\bf s}$ can always 
be chosen to vanish, and therefore never needs to be 
considered~\cite{martin}. While the first part of this stance is fairly 
convincing, the final conclusion excludes a whole category of hydrodynamic 
phenomena. These include especially the Einstein-de Haas effect, in which a 
stationary magnetizable body starts to rotate when the external magnetic 
field is turned off~\cite{LL8}: The vanishing velocity {\boldmath$v$} of a 
stationary body also compels the momentum density $\mbox{\boldmath$g$} 
=\varrho\,\mbox{\boldmath$v$}$ and the orbital angular momentum density 
$\mbox{\boldmath$x$}\times\mbox{\boldmath$g$}$ to be zero. But the total 
angular momentum ${\bf L}$ may still be finite in the presence of a magnetic 
field $\mbox{\boldmath${\cal H}$}$ if there is a finite intrinsic angular 
momentum ${\bf s}\sim\mbox{\boldmath${\cal H}$}$. If 
$\mbox{\boldmath${\cal H}$}$ is turned off, ${\bf s}$ vanishes, yet ${\bf L}$ 
must remain constant. Hence the system starts rotating, as observed, to 
compensate. 

The analogous effect for the booster would be given by aligning the 
microscopic mass dipoles with an external electric field, ie ${\bf k}\sim 
\mbox{\boldmath${\cal E}$}$. Turning the field off gets rid of the alignment 
and kills ${\bf k}$. The orbital part $M{\bf R}_0$ compensates by displacing 
the crystal. 

The proper way to account for this type of effects lies in the middle 
ground between the above two extreme points of view --- taking ${\bf s}$ as 
either completely independent or utterly negligible.  As discussed in 
section~\ref{4}, we should in fact accept ${\bf s}$ as an thermodynamic 
variable, but not a hydrodynamic one (the precise meaning of which will 
become clear there). The close analogy between ${\bf s}$ and ${\bf k}$ 
makes it easy to treat the latter along the same line. One of the results, 
based on fairly general thermodynamic considerations, is that the booster's 
analogue of the Einstein-de Haas effect does not exist.

{\bf v)} Finally, a truly serendipitous result: the collective, hydrodynamic 
behavior of systems that spontaneously break Galilean symmetries. This 
subject has until now eluded clarification because the relevant conserved 
quantity, the  booster, has been neglected. Again, the analogy between the 
angular momentum and the booster comes in handy here, as broken rotational 
symmetry is a well understood concept~\cite{martin,dG}. The order parameter 
${\bf u}$ of the broken Galilean symmetry obeys the equation of  motion, 
$\partial_t{\bf u}+\mbox{\boldmath$\nabla$}\mu=0$. Despite the obvious 
similarity to the Josephson equation~\cite{Kh}, the Goldstone modes of 
broken Galilean symmetries are not propagating second sound modes, but 
resemble the orbital waves of nematic liquid crystals. 

Unfortunately, we have nothing to say concerning possible microscopic 
mechanisms that realize broken Galilean symmetry.  The system that 
spontaneously breaks this symmetry needs to display indifference with 
respect to the inertial system it chooses while being able to sustain a 
velocity gradient in equilibrium. And the relevant order parameter ${\bf 
u}$ is a velocity difference.  Such as system is admittedly hard to 
conceive. 

\section{Frame-Independent Thermodynamics}\label{2}
Let us begin by deducing the thermodynamic equilibrium conditions.
Maximizing the entropy, or equivalently, minimizing the energy, while 
holding constant the entropy and the conserved quantities, including 
especially the booster {\bf B}, we have

\begin{eqnarray}\label{v1}
\delta\,\mbox{\Large[}
&&\,{\textstyle\int}{\rm d}^3\!x\,\varepsilon - \hat T\,
  {\textstyle\int}{\rm d}^3\!x\,s - \hat\mu\,{\textstyle\int}{\rm d}^3\!x\,
  \varrho - \hat{\bf V}\!\cdot\!{\textstyle\int}{\rm d}^3\!x\,
  \mbox{\boldmath$g$}\nonumber\\ 
&&\quad - \hat{\bf\Omega}\!\cdot\!{\textstyle\int}{\rm d}^3\!x\,
(\mbox{\boldmath$x$}\!\times\!\mbox{\boldmath$g$}) - \hat{\bf A}\!\cdot\!
{\textstyle\int}{\rm d}^3\!x\,(\varrho\,\mbox{\boldmath$x$}- 
\mbox{\boldmath$g$}\,t)\,\mbox{\Large]}=0, 
\end{eqnarray}
with $\hat T$, $\hat\mu$, $\hat{\bf V}$, $\hat{\bf\Omega}$, and $\hat{\bf A}$
being constant Lagrange para\-me\-ters. Employing the local Gibbs relation, 

\begin{equation}\label{v3}
d\varepsilon=T\,ds+\mu\,d\varrho
+\mbox{\boldmath$v$}\!\cdot\!d\mbox{\boldmath$g$},
\end{equation} 
this can be written as

\begin{eqnarray}\label{v5}
{\textstyle\int}{\rm d}^3\!x\,\mbox{\Large[}\,
&&(T-\hat T)\,\delta s + (\mu-\hat\mu-\hat{\bf A}\!\cdot\!\mbox{\boldmath$x$})
  \,\delta\varrho\nonumber\\ 
&&\ + (\mbox{\boldmath$v$}-\hat{\bf V}-\hat{\bf\Omega}\!\times\!
    \mbox{\boldmath$x$}+\hat{\bf A}\,t)\!\cdot\!\delta\mbox{\boldmath$g$}
    \,\mbox{\Large]}=0.
\end{eqnarray}
From this follow the Euler-Lagrange equations

\begin{equation}
T=\hat T,\quad \mu=\hat\mu+\hat{\bf A}\cdot\mbox{\boldmath$x$},\quad 
\mbox{\boldmath$v$}=\hat{\bf V}+\hat{\bf\Omega}\!\times\!\mbox{\boldmath$x$}
-\hat{\bf A}\,t.
\label{v}\end{equation}
Comparing the last of Eqs(10) to the velocity field of Eq(\ref{rig}), we find complete 
agreement with 

\begin{equation}\label{v2}
 \hat{\bf A}={\bf\Omega\times\dot R},\quad\hat{\bf\Omega}= 
{\bf\Omega},\quad\hat{\bf V}={\bf\dot R- \Omega\times R}_0.
\end{equation}
Excluding the conservation of the booster is equivalent to setting 
$\hat{\bf A}={\bf\Omega\times\dot R}=0$, cf Eq(\ref{v1}). This 
leads to the results of Landau and Lifshitz, mentioned in the 
introduction, correct for frames in which ${\bf\Omega\,\|\,\dot R}$.
 
In a more deliberate approach, the input of Eq(\ref{rig}) is not necessary: 
Varying also the time and space coordinates while minimizing the energy 
yields two additional terms, 
\begin{equation}\label{v4}
{\textstyle\int}{\rm d}^3\!x\,\mbox{\Large[}\dots+(\hat{\bf\Omega}\!\times\!
\mbox{\boldmath$g$}-\hat{\bf A}\varrho)\!\cdot\!\delta\mbox{\boldmath$x$}+
(\hat{\bf A}\!\cdot\mbox{\boldmath$g$})\,\delta t\,\mbox{\Large]}=0.
\end{equation} 
Taking uniform transformations in time and space, $\delta t,\,\delta 
\mbox{\boldmath$x$}=const$, as independent variations, we conclude   
$\hat{\bf A}={\bf\Omega\times\dot R}$ and (redundantly) $\hat{\bf 
A}\cdot{\bf\dot R=0}$. 

Locally expressed, Eqs(\ref{v}) become 
\begin{equation}
\mbox{\boldmath$\nabla$}T=0,\quad\partial_i v_j+\partial_j v_i=0,\quad 
\partial_t\mbox{\boldmath$v$}+ \mbox{\boldmath$\nabla$}\mu=0,
\label{loc}\end{equation}
which are useful for the off-equilibrium considerations below. There are 
three corollary conclusions to draw:

(i) Integrating Eq(\ref{v3}) over the system's constant volume while 
heeding Eqs(\ref{v}), we obtain the Galilean covariant version of the 
extensive, basic formula of the thermodynamics, \begin{equation}
dE=\hat TdS+\hat\mu\,dM+\hat{\bf V}\!\cdot d{\bf G}+ 
   \hat{\bf\Omega}\cdot d{\bf L}+\hat{\bf A}\!\cdot d{\bf B}.
\label{ex}\end{equation}
If the volume is allowed to vary, there is an additional term, 
$-\oint\!{\rm d}^2x\,P\,du$, with $P\equiv\mu\varrho+ Ts+\mbox{\boldmath$v$} 
\cdot\mbox{\boldmath$g$}- \varepsilon$ and $du$ denoting the displacement of 
the surface, along the surface normal and at the area element ${\rm d}^2x$. 
The unusual form is related to the fact that $P$ is not a constant, and the 
energy change depends on where the volume change takes place. For $P=const$, 
this term reduces to the usual form $-P\oint\!{\rm d}^2x\,du=-P\,dV$. The 
last three terms of Eq(\ref{ex}), ie the kinetic part of the energy,  can be 
integrated (for a sphere) to become 
\begin{equation}\label{ex2}
E_{kin}={\bf G}^2/2M+({\bf L}-{\bf R}_0\!\times\!{\bf G})^2/2\Theta,
\end{equation}
as it should. $\Theta$ is the moment of inertia in the center of mass frame.

(ii) Despite the manifest frame dependence of $\mu$ and {\boldmath $v$} 
in Eq(\ref{v}), most quantities are properly frame independent. Take 
eg the density distribution under rotation: The chemical potential 
of the local rest frame, $\mu_0=\mu+\frac{1}{2}v^2$, is a 
function of $T$ and $\varrho$, hence $\mbox{\boldmath$\nabla$}\mu_0 
=(\partial\mu_0/\partial\varrho)\mbox{\boldmath$\nabla$}\varrho$ for 
constant temperatures. On the other hand, we deduce from Eqs(\ref{v}, 
\ref{v2}) that 

\begin{equation}\label{ex1}
\mbox{\boldmath$\nabla$}\mu_0= \mbox{\boldmath$\nabla$}(\mu+
\textstyle\frac{1}{2}v^2)= 
\textstyle\frac{1}{2}\mbox{\boldmath$\nabla$}[{\bf\Omega} 
\times(\mbox{\boldmath$x$}-{\bf R})]^2
\end{equation}
depends only on the velocity in the center of mass frame, 
and not on ${\bf\dot R}$, the center of mass velocity of a given frame. So 
$\mbox{\boldmath$\nabla$}\mu_0$, and therefore 
$\mbox{\boldmath$\nabla$}\varrho$, remain unchanged under a Galilean boost. 

(iii) The Navier-Stokes equation  is 
$d\mbox{\boldmath$v$}/dt+\mbox{\boldmath$\nabla$}P/\varrho=0$
for $\partial_i v_j+\partial_j v_i=0$, and 
$d\mbox{\boldmath$v$}/dt+\mbox{\boldmath$\nabla$}\mu_0=0$ 
if in addition $\mbox{\boldmath$\nabla$}T=0$. Rewriting $d\mbox{\boldmath$v$}/dt 
\equiv\partial_t\mbox{\boldmath$v$} 
 +(\mbox{\boldmath$v$}\cdot\!\mbox{\boldmath$\nabla$})\mbox{\boldmath$v$} 
= \partial_t\mbox{\boldmath$v$}-\frac{1}{2}\mbox{\boldmath$\nabla$}v^2$, it 
finally reduces to the third of Eqs(\ref{loc}), 
$\partial_t\mbox{\boldmath$v$}+ \mbox{\boldmath$\nabla$}\mu =0$.

\section{The Dissipative Mass Current}\label{3}
The inclusion of the booster is clearly important for static, 
thermodynamic considerations, but it is equally relevant for 
off-equilibrium, dynamic situations: First, the proof for 
$\mbox{\boldmath$\jmath$}=\mbox{\boldmath$g$}$, or that no dissipative mass 
current is allowed. As this proof follows closely the one that 
deduces the symmetry of the stress tensor from 
local conservation of angular momentum, it aids comprehension to present 
both simultaneously. Rewriting the continuity equations for mass and momentum,
\begin{equation}\label{lc3}
\partial_t\varrho+\partial_i\jmath_i=0,\quad 
\partial_tg_i+\partial_j\Pi_{ij}=0,
\end{equation}as
\begin{eqnarray}
\partial_t\ell_i+\partial_j(\epsilon_{ikm}x_k\Pi_{mj}) 
=\epsilon_{ijk}\Pi_{kj}\,,\\ 
\partial_tb_i+\partial_j(x_i\jmath_j-\Pi_{ij}t)=\jmath_i-g_i\,,
\label{lc2}\end{eqnarray}
where $\mbox{\boldmath$\ell$}\equiv\mbox{\boldmath$x$}\!\times\!
\mbox{\boldmath$g$}$ and ${\bf b}\equiv\varrho\, 
\mbox{\boldmath$x$} -\mbox{\boldmath$g$}\,t$, we immediately see that 
$\epsilon_{ijk}\Pi_{kj}$ and $\mbox{\boldmath$\jmath$}-\mbox{\boldmath$g$}$ 
have to vanish for the angular momentum and the booster to be locally 
conserved. 

Now, one may argue that it would be quite enough if these two expressions 
can be written as divergences of some currents, 
$\epsilon_{ijk}\Pi_{kj}=\partial_k\sigma_{ik}$ and 
$\jmath_i-g_i=\partial_kJ_{ik}$, so they need not vanish. But 
this neglects the following two points: First, qualitatively, 
$\mbox{\boldmath$\ell$}$ and {\bf b} contain the reference to the 
origin of the coordinate, $\mbox{\boldmath$g$}$ and $\varrho$ do not. 
Consider a small volume element far away from the origin, at distance ${\bf 
R}$ and $t=0$, then $\mbox{\boldmath$\ell$}$ and ${\bf b}= 
\varrho\mbox{\boldmath$x$}$ will scale with ${\bf R}$. Now, as we can change 
${\bf R}$ by simply relocating the origin, without altering any of the 
physics, $\partial_t\mbox{\boldmath$\ell$}$, $\partial_t 
{\bf b}$ and their respective fluxes must also scale with 
${\bf R}$. On the other hand, because $\Pi_{ij}$ and 
$\mbox{\boldmath$\jmath$}$ (as fluxes of $\mbox{\boldmath$g$}$ and 
$\varrho$) do not, neither do $\epsilon_{ijk}\Pi_{kj}=\partial_k\sigma_{ik}$ 
and $\jmath_i-g_i=\partial_kJ_{ik}$, which therefore must vanish. 

Second, the quantitative argument. Being currents of hydrodynamic 
variables, $\Pi_{ij}$ and $\jmath_i$ are themselves functions of these 
variables, and of their spatial derivatives. To lowest order, with no spatial 
derivatives, $\Pi_{ij}$ and $\jmath_i$ cannot possibly be written as divergences 
of some currents.  In 
the next order, the terms are mostly dissipative. Taking 
$\epsilon_{ijk}\Pi_{kj}$, $\mbox{\boldmath$\jmath$}-\mbox{\boldmath$g$}$ as 
finite, we find the entropy production to be given as~\cite{Vol,GM}  

\begin{equation}
{\cal R}=\dots+ \epsilon_{ijk}\Pi_{jk}\Omega_i-(\mbox{\boldmath$\jmath$}
  -\mbox{\boldmath$g$})\!\cdot\!\mbox{\boldmath$\nabla$}\mu, 
\quad {\bf\Omega}\equiv{\textstyle\frac{1}{2}}
      \mbox{\boldmath$\nabla$}\!\times\!\mbox{\boldmath$v$}.
\label{R1}\end{equation}
According to the rules of irreversible thermodynamics, if the entropy 
production is given as a sum of products, ${\cal R}=\Sigma X_iY_i$, we may 
take the fluxes $X_i$ as proportional to the thermodynamic forces 
$Y_i$~\cite{GM}, hence 

\begin{equation}\label{R2}
\epsilon_{ijk}\Pi_{jk} =\zeta_{\scriptscriptstyle(1)} 
\,\Omega_i, \quad
\mbox{\boldmath$\jmath$}-\mbox{\boldmath$g$}=-\zeta_{\scriptscriptstyle(2)} 
\,\mbox{\boldmath$\nabla$}\mu, 
\end{equation} 
where $\zeta_{\scriptscriptstyle(1)}$ and 
$\zeta_{\scriptscriptstyle(2)}$ are transport coefficients, similar to the 
viscosity. (More accurately, the $X_i$ are a linear combination of the $Y_i$, 
so off-diagonal, cross terms are also possible. However, these terms always 
vanish if the diagonal ones do, so they need no extra consideration.) 
Eqs(\ref{R2}) lead to the entropy production 

\begin{equation}\label{R3}
{\cal R}=\dots+ \zeta_{\scriptscriptstyle(1)} \,{\bf\Omega}^2 
+\zeta_{\scriptscriptstyle(2)} \,(\mbox{\boldmath$\nabla$}\mu)^2
\end{equation}
which, however, contradicts Eqs(\ref{loc}): ${\bf\Omega}$ and 
$\mbox{\boldmath$\nabla$} \mu$ need not vanish even if all the three fields 
there do, in which case equilibrium reigns with ${\cal R}\equiv0$. Therefore 
$\zeta_{\scriptscriptstyle(1)}$, $\zeta_{\scriptscriptstyle(2)}$ are zero, 
and $\epsilon_{ijk}\Pi_{kj}$, $\mbox{\boldmath$\jmath$}-\mbox{\boldmath$g$}$ 
vanish.

\section{Intrinsic Booster-Density}\label{4}
Allowing now intrinsic contributions, as in Eqs(\ref{intr}, \ref{intr2}), 
the energy density depends on two additional variables, ${\bf s}$ and $\bf 
k$, or  

\begin{equation}
d\varepsilon=Tds+\mu\,d\varrho+\mbox{\boldmath$v$}\cdot d\mbox{\boldmath$g$} 
+\mbox{\boldmath$\omega$}\cdot d{\bf s}+\mbox{\boldmath$a$}\cdot d{\bf k}\,. 
\label{e}\end{equation}
Barring an instability, the energy $\varepsilon$ is minimal for ${\bf s, 
k}=0$. Therefore, an expansion around the minimal value, 
$\varepsilon_{min}$, would usually yield a quadratic  dependence, 
$\varepsilon=\varepsilon_{min}+ \textstyle\frac{1}{2}
(\gamma_{\scriptscriptstyle (1)}{\bf s}^2 
+\gamma_{\scriptscriptstyle (2)}{\bf k}^2)$. (Isotropy is assumed, otherwise 
$\gamma_{\scriptscriptstyle(1)}$, $\gamma_{\scriptscriptstyle(2)}$ are 
tensors rather than scalars.) Being derivatives, {\boldmath$\omega$} and 
{\bf a} assume the form 

\begin{equation}\label{e2}
\mbox{\boldmath$\omega$}= \gamma_{\scriptscriptstyle (1)}{\bf s},
\quad\mbox{\boldmath$a$}=\gamma_{\scriptscriptstyle (2)}{\bf k}.
\end{equation}
If the magnetic field is included, $d\varepsilon=\dots
+\mbox{\boldmath${\cal H}$}\cdot d\mbox{\boldmath${\cal B}$}$, the energy 
is minimal for  ${\bf s, k}, \mbox{\boldmath${\cal B}$}=0$, and the same 
expansion (again for an isotropic medium) yields 

\begin{equation}\label{e3}
\mbox{\boldmath$\omega$}= \gamma_{\scriptscriptstyle (1)}{\bf s} + 
\gamma_{\scriptscriptstyle (3)}\mbox{\boldmath${\cal B}$}.
\end{equation}
It is this cross dependency that leads to the Einstein-de Haas effect 
discussed in the introduction~\cite{LL8}. Similarly, an electric field leads 
to 

\begin{equation}\label{e4}
\mbox{\boldmath$a$}= \gamma_{\scriptscriptstyle (2)}{\bf k} + 
\gamma_{\scriptscriptstyle (4)}\mbox{\boldmath${\cal D}$}.
\end{equation}
Because $\mbox{\boldmath$\ell$}+{\bf s}$ and ${\bf b}+{\bf 
k}$ are locally conserved quantities and satisfy continuity equations, the 
new variables obey the equations of motion, 

\begin{equation}
\partial_ts_i+\partial_j\lambda_{ij}=\epsilon_{ijk}\Pi_{jk},\quad 
\partial_tk_i+\partial_j\gamma_{ij}=g_i-\jmath_i,
\label{s+c}\end{equation} 
where $\lambda_{ij}$ and $\gamma_{ij}$ are the respective currents, 
functions of the local hydrodynamic variables, the explicit form of which we 
shall not derive here. The source terms on the right hand side cancel those 
of Eqs(\ref{lc2}), and ensure that $\mbox{\boldmath$\ell$}+{\bf s}$ and 
${\bf b}+{\bf k}$ satisfy continuity equations. (The 
vanishing of $\epsilon_{ijk}\Pi_{jk}$ and 
$\mbox{\boldmath$g$}-\mbox{\boldmath$\jmath$}$ was deduced in the last 
section as a direct consequence of $\mbox{\boldmath$\ell$}$ and 
${\bf b}$ being locally conserved quantities. As this is no longer 
true, we have no prior knowledge of whether $\epsilon_{ijk}\Pi_{jk}$ and 
$\mbox{\boldmath$g$}- \mbox{\boldmath$\jmath$}$ are finite.) Going through 
the same hydrodynamic procedure as that leading to Eq(\ref{R1}), we find the 
modified entropy production, 

\begin{equation}
{\cal R}=\dots+\epsilon_{ijk}\Pi_{jk}(\Omega_i-\omega_i)
-(\mbox{\boldmath$\jmath$}-\mbox{\boldmath$g$})\!\cdot\!
(\mbox{\boldmath$\nabla$}\mu-\mbox{\boldmath$a$}),
\label{lax1}\end{equation}
and conclude as before, 
\begin{equation}
\epsilon_{ijk}\Pi_{jk}=\zeta_{\scriptscriptstyle 
(1)}(\Omega_i-\omega_i),\quad 
\mbox{\boldmath$\jmath$}-\mbox{\boldmath$g$}=-\zeta_{\scriptscriptstyle 
(2)}(\mbox{\boldmath$\nabla$}\mu-\mbox{\boldmath$a$}), 
\label{lax}\end{equation}
If the two coefficients were exceedingly large, 
$\zeta_{\scriptscriptstyle (1)}$, $\zeta_{\scriptscriptstyle (2)}\to\infty$, 
we may divide each of Eqs(\ref{s+c}) by the respective coefficient and find 
the equations substituted by the two constrains, 

\begin{equation}
\mbox{\boldmath$\omega$}={\bf\Omega}\,, \quad 
\mbox{\boldmath$a$}=\mbox{\boldmath$\nabla$}\mu\,.
\label{con}\end{equation}
The first implies instantaneous establishment of partial equilibrium by 
local exchanges of angular momentum, between the reservoirs 
$\mbox{\boldmath$\ell$}$ and ${\bf s}$; the second implies partial 
equilibrium by local exchanges of booster, between 
${\bf b}$ and ${\bf k}$. So instead of an independent 
dynamics, ${\bf s}$ and ${\bf k}$ assume the appropriate values satisfying 
Eqs(\ref{con}), instantaneously during hydrodynamic processes.   

For generic circumstances, this is indeed the case. There are two terms in 
Eqs(\ref{lax}), $$\zeta_{\scriptscriptstyle 
(1)}\mbox{\boldmath$\omega$}=\zeta_{\scriptscriptstyle (1)}
\gamma_{\scriptscriptstyle (1)}{\bf s}\equiv {\bf s}/
\tau_{\scriptscriptstyle (1)},\quad \zeta_{\scriptscriptstyle (2)}
 \mbox{\boldmath$a$}=\zeta_{\scriptscriptstyle (2)}
\gamma_{\scriptscriptstyle (2)}{\bf k} \equiv{\bf k}/ 
\tau_{\scriptscriptstyle (2)},$$ which show that both Eqs(\ref{s+c}) 
are relaxation equations, and that Eqs(\ref{con}) hold after the time 
$\tau_{\scriptscriptstyle (1)}$ and $\tau_{\scriptscriptstyle (2)}$, 
respectively.  Generally speaking, these characteristic times, after which 
local equilibrium is established, are microscopic in nature and much faster 
than typical hydrodynamic time scales. And taking these times as negligibly 
small is equivalent to the limit $\zeta_{\scriptscriptstyle (1)}$, 
$\zeta_{\scriptscriptstyle (2)}\to\infty$, and equivalent to substituting Eqs(\ref{con}) 
for the equations of motion~(\ref{s+c}). Note, however, that this dynamic 
dependency does not at all mean ${\bf s}$ and ${\bf k}$ can not remain 
full-fledged thermodynamic variables, as 
Eqs(\ref{e},\ref{e2},\ref{e3},\ref{e4}) clearly remain valid. 

From a more elevated point of view, we understand that the universality and 
simplicity of hydrodynamic theories are achieved by eliminating all relaxing 
variables, and expunging all relaxation equations. As a result, the 
hydrodynamic theory is confined to frequencies much less than all relaxation 
rates $1/\tau_{\scriptscriptstyle (i)}$. 

(A possible source of confusion is the seemingly odd fact that the 
transport coefficients $\zeta_{\scriptscriptstyle (1)}$ and 
$\zeta_{\scriptscriptstyle (2)}$ were deduced to be zero in the last 
section, yet argued to be diverging here. The explanation lies in the 
difference between the calibers of the arguments: In the last section, the 
coefficients were not negligibly small. Rather, they had to vanish 
identically to not contradict thermodynamics. Here, the coefficients are 
finite, but well approximated by $\infty$ when compared to hydrodynamic time 
scales. In other words, if they are finite, they are very large -- and there is 
no contradiction.)

One problem remains: The two terms $\epsilon_{ijk}\Pi_{jk}$ and 
$\mbox{\boldmath$\jmath$}-\mbox{\boldmath$g$}$ are now indeterminate, as 
they are both products of an infinite and a vanishing factor, cf 
Eqs(\ref{lax}).  Nevertheless, they need to be known before we may make use 
of the continuity equations for mass and momentum, Eqs(\ref{lc3}). A clever 
way to accomplish this is to go over to the new densities~\cite{martin}, 

\begin{equation}
\mbox{\boldmath$\tilde g$}=\mbox{\boldmath$g$}+{\textstyle\frac{1}{2}}
(\mbox{\boldmath$\nabla$}\!\times{\bf s})\,,\quad 
\tilde\varrho=\varrho-\mbox{\boldmath$\nabla$}\!\cdot{\bf k}\,. 
\label{nd}\end{equation}
They are obtained by starting from Eq(\ref{e}), with Eqs(\ref{con}) 
incorporated, 

\begin{equation}\label{nd1}
d\varepsilon=\dots\mbox{\boldmath$v$}\cdot 
d\mbox{\boldmath$g$}+{\bf\Omega}\cdot d{\bf s}+ 
\mu\,d\varrho+\mbox{\boldmath$\nabla$}\mu\cdot d{\bf k}. \end{equation}
This is partially integrated to yield 

\begin{equation}\label{nd2}
d\varepsilon=\dots + \mbox{\boldmath$v$}\cdot d\mbox{\boldmath$\tilde g$}+ 
\mu\,d\tilde\varrho,  
\end{equation}
which shows first of all that the energy density is well accounted for by 
the new densities. Putting the surface slightly beyond the system's volume, 
we also have 

\begin{equation}\label{nd3}
{\textstyle\int}{\rm d}^3\!x\,\mbox{\boldmath$\tilde 
g$}={\textstyle\int}{\rm d}^3\!x\,\mbox{\boldmath$g$},\quad 
{\textstyle\int}{\rm d}^3\!x\,\tilde
 \varrho={\textstyle\int}{\rm d}^3\!x\,\varrho,
\end{equation}
as the respective second term of Eq(\ref{nd}) are surface terms. So we 
conclude that $\mbox{\boldmath$\tilde g$}$ and $\tilde\varrho$ are also 
valid densities for momentum and mass. Most crucially, because of 

\begin{equation}\label{nd4}
\ {\textstyle\int}{\rm 
d}^3\!x\,(\mbox{\boldmath$x$}\times\mbox{\boldmath$\tilde g$})=
{\textstyle\int}{\rm 
d}^3\!x\,(\mbox{\boldmath$x$}\times\mbox{\boldmath$g$}+{\bf s}), 
\end{equation}
and 
\begin{equation}\label{nd5}
{\textstyle\int}{\rm d}^3\!x\,(\tilde\varrho\,\mbox{\boldmath$x$} 
-\mbox{\boldmath$\tilde g$}\,t)= {\textstyle\int}{\rm d}^3\!x\,(\varrho\,
\mbox{\boldmath$x$} -\mbox{\boldmath$g$}\,t+{\bf k}), 
\end{equation}
these new densities clearly absorb the intrinsic contributions, ${\bf s}$, 
${\bf k}$, and therefore have fluxes for which 

\begin{equation}\label{nd6}
\epsilon_{ijk}\tilde\Pi_{jk}=0,\quad 
\mbox{\boldmath$\tilde\jmath$}-\mbox{\boldmath$\tilde g$}=0, 
\end{equation}
hold. So it is indeed a good idea to take $\tilde\varrho$ and 
{\boldmath$\tilde g$} as the hydrodynamic variables. (But it is now also 
possible to return to the original variables: Given $\tilde\Pi_{ij}$ and 
{\boldmath$\tilde\jmath$}, it is easy to obtain $\Pi_{ij}= \tilde\Pi_{ij}+ 
\frac{1}{2}\epsilon_{ijk}\partial_t s_k$, $\mbox{\boldmath$\jmath$} = 
\mbox{\boldmath$\tilde\jmath$}- \partial_t{\bf k}$.)

All this does not mean that no experimental consequences are 
related to ${\bf s}$ and ${\bf k}$, as claimed in~\cite{martin}. For 
instance, the constitutive relations, 
$\mbox{\boldmath$g$}=\varrho\mbox{\boldmath$v$}$ and Eq(\ref{e3}), or 

\begin{equation}\label{nd7}
\mbox{\boldmath$\tilde g$}=\varrho\mbox{\boldmath$v$}+{\textstyle\frac{1}{2}}
\mbox{\boldmath$\nabla$}\!\times{\bf s}(\mbox{\boldmath$\Omega$},
\mbox{\boldmath${\cal B}$}),
\end{equation}
can not be without consequences in the Navier-Stokes equation, 
$\partial_t\tilde g_i=\tilde\Pi_{ij}$:  A temporally varying but spatially 
homogeneous $\mbox{\boldmath${\cal B}$}$-field will lead to an acceleration,   
$\partial_t(\varrho\,\mbox{\boldmath$v$})=-\partial_t[\mbox{\boldmath$\nabla$}
\!\times{\bf s}(\mbox{\boldmath${\cal B}$})/2]$, at the system's 
outer rim, the only place where $\mbox{\boldmath$\nabla$}\!\times{\bf s}$ 
is nonzero. The resultant rotary motion there then propagates into the bulk, 
elastically in a solid and viscously in a liquid. Clearly, this is a 
temporally resolved, hydrodynamic description of the Einstein-de Haas 
effect. 

Other thermodynamic cross dependencies of $\bf s$ will lead to similar 
behavior. The diagonal term all by itself, however, while necessary for 
maintaining the thermodynamic stability, is probably always negligibly 
small, as $$\mbox{\boldmath$\nabla$}\!\times{\bf s}/2= 
\mbox{\boldmath$\nabla$}\!\times\mbox{\boldmath$\Omega$}/ 
2\gamma_{\scriptscriptstyle (1)}$$  is of the same order as the viscous 
terms. 

In spite of the one-to-one analogy, $$\partial_t\tilde\varrho 
=\partial_t[\varrho- \mbox{\boldmath$\nabla$}\!\cdot{\bf k}]$$  instead of 
Eq(\ref{nd7}), there is one crucial difference between ${\bf s}$ and ${\bf 
k}$: The quantity $$\partial\varepsilon/\partial{\bf 
k}=\mbox{\boldmath$\nabla$}\mu={\bf\Omega\times\dot R}$$ depends on the 
inertial frame. As a result, thermodynamic cross dependencies are ruled out 
for $\bf k$. To understand why, consider 
${\bf k}(\mbox{\boldmath${\cal E}$})$, the electric analog 
of ${\bf s}(\mbox{\boldmath${\cal B}$})$. Due to the Maxwell relation, 

\begin{equation}\label{x1}
\partial k_j/\partial {\cal E}_i=\partial {\cal D}_i/\partial(\partial_j\mu), 
\end{equation}
that follows from the energy $d\varepsilon=\dots {\cal E}_id{\cal D}_i+ 
(\partial_j\mu)dk_j$, $\ {\bf k}$'s dependence on $\mbox{\boldmath${\cal E}$}$
implies $\mbox{\boldmath${\cal D}$}$'s on ${\bf\mbox{\boldmath$\nabla$}\mu}$, 
a quantity that can be made to vanish by choosing an appropriate frame. 
In the absence of a magnetic field, 
$\mbox{\boldmath${\cal D}$}$ is essentially frame independent, 
$\mbox{\boldmath${\cal D}$}(\mbox{\boldmath$\nabla$}\mu)=
\mbox{\boldmath${\cal D}$}(0)$, or 
$\partial {\cal D}_i/\partial(\partial_j\mu)=0$. Hence 
$\partial k_j/\partial {\cal E}_i=0$. 

The diagonal term $\gamma_{\scriptscriptstyle (2)}{\bf 
k}=\mbox{\boldmath$\nabla$}\mu= {\bf\Omega\times\dot R}$ remains, but the 
associated hydrodynamic effects, 
$\partial_t\mbox{\boldmath$\nabla$}\cdot{\bf k}\sim(\partial\mu/ 
\partial\varrho)\partial_t\nabla^2\varrho$ are again an order smaller than 
the dissipative terms. They may probably always be ignored. Nevertheless, 
understanding the treatment of a finite ${\bf k}$ is useful, both for its 
own sake and  for the consideration  below of broken Galilean symmetry, 
(although these system do not necessarily have a finite ${\bf k}$.)

\section{Spontaneously Broken Galilean Symmetry}\label{5}
Thermo- and hydrodynamic theories account for the generic behavior of 
macroscopic systems that results from conservation laws and broken 
symmetries. Although the specificity of its predictive power is, compared 
to the general input, frequently amazing, some subjects are simply beyond 
its reach. One example is the question concerning possible microscopic 
mechanisms which lead to broken Galilean symmetries, or in fact 
whether it at all exists. Fortunately, we do not need this knowledge to 
give a full account of the collective behavior of a system that breaks this 
symmetry, and hereby filling the macroscopic half of a large gap in one of 
our more basic concepts. 
 
Starting with a system that breaks rotational invariance in all three 
directions, (say biaxial nematics or the B-phase of superfluid $^3$He 
\cite{bn},) we take the infinitesimal rotation angle $d\theta_i$ as the 
order parameter, with the energy depending on its gradient, 

\begin{equation}\label{x2}
d\varepsilon=\dots+ \Omega_ids_i + \phi_{ij}d(\partial_j\theta_i).
\end{equation}
An expansion of the energy $\varepsilon$ leads to 
$\phi_{kl}\sim\partial_j\theta_i$. The equations of motion in the rest frame 
possess the usual Hamiltonian form, 

\begin{equation}
\partial_t\theta_i= \delta\varepsilon/ \delta s_i = 
\Omega_i,\quad  
\partial_t s_i = -\delta\varepsilon/ \delta\theta_i = 
\partial_j\phi_{ij}.
\label{brokenR}\end{equation} 
Eliminating $\partial_t s_i$ as outlined in the last section, the 
continuity equation for the momentum gets modified:  $$\partial_t\tilde g_i= 
\partial_tg_i+ \frac{1}{2}\epsilon_{ijk}\partial_j\partial_ts_k 
=\frac{1}{2}\epsilon_{ijk}\partial_j\partial_m\phi_{km}+\dots$$ 
which leads to three pairs of orbital waves, of the form $\omega\sim q^2$, 
with complex coefficients~\cite{bn}. (A slight rearrangement is needed to 
render the seemingly antisymmetric stress tensor symmetric.) 

The order parameter of broken Galilean symmetry is a velocity field, $u_i$, 
with the energy depending on its gradient, 

\begin{equation}\label{x3}
d\varepsilon= \dots+(\partial_i\mu)dk_i+J_{ij}d(\partial_ju_i). 
\end{equation}
An expansion leads to 
$J_{kl}\sim\partial_ju_i$.  
The equations of motion in the rest frame again possess the usual 
Hamiltonian form, 

\begin{equation}
\partial_t u_i= -\delta\varepsilon/ \delta k_i = -\partial_i\mu,\quad 
\partial_tk_i =\delta\varepsilon/\delta u_i=-\partial_jJ_{ij}. 
\label{brokenG}\end{equation}
Eliminating $\partial_tk_i$ as outlined in the last section, the continuity 
equation for the mass density gets modified:  
$$\partial_t\tilde\varrho= \partial_t\varrho-\partial_i\partial_t k_i = 
-\partial_i(\varrho\,v_i- \partial_jJ_{ij}),$$ 
where the total mass current equals the momentum density. Including 
nonlinear terms and the diagonal dissipative one, we have (dropping 
$\tilde{\phantom a}$), 

\begin{eqnarray}
\partial_t u_i+(v_k\partial_k)u_i+\partial_i\mu-\zeta\,\partial_jJ_{ij}=0,\\
\partial_t\varrho+\partial_ig_i=0,\quad g_i=\varrho\,v_i-\partial_j J_{ij}.
\label{bG}\end{eqnarray}
In conjunction with the rest of the hydrodynamic equations that essentially 
retain their form from the isotropic liquid, we find the sound mode to be 
unchanged, while three additional modes, of coupled temperature and 
$u$-motion, are (as in the orbital case) of the form $\omega\sim q^2$, with 
the coefficients being complex.

\end{multicols}
\end{document}